\documentclass[twocolumn,letter]{jpsj3}
\usepackage{txfonts}
\usepackage{graphicx}% Include figure files
\usepackage{dcolumn}% Align table columns on decimal point
\usepackage{bm}% bold math
\usepackage[usenames,dvipsnames]{xcolor}
\usepackage{ulem}
\usepackage{braket}
\usepackage{amsmath}
\usepackage{tabularx}

\title{Field-Direction Sensitive Skyrmion Crystals in Cubic Chiral Systems: \\ Implication to $4f$-Electron Compound EuPtSi}

\author{Satoru Hayami and Ryota Yambe}
\inst{
Department of Applied Physics, The University of Tokyo, Tokyo 113-8656, Japan
 }

\abst{
We theoretically study the stability of magnetic skyrmion crystals (SkXs) in chiral cubic antiferromagnets with the EuPtSi hosting unconventional nanometric SkXs in mind. 
By performing numerical simulations for a minimal effective spin model with the long-range Dzyaloshinskii-Moriya interaction and the multiple-spin interaction arising from itinerant nature of electrons, we clarify two key ingredients for the emergence of the SkXs in EuPtSi: 
One is the low-symmetric ordering vectors that lead to the field-direction sensitive SkXs in an external magnetic field, while the other is the synergy between spin-charge and spin-orbit couplings in itinerant magnets that results in the nanometric SkXs. 
The present study will not only provide the microscopic mechanism of the SkXs in EuPtSi but also guide for a search of further materials hosting short-period SkXs.
}

\begin{document}
\maketitle

A topological spin texture that accompanies a spin scalar chirality by a triple spin product, $\bm{S}_i \cdot (\bm{S}_j \times \bm{S}_k)$, has received much attention in recent years, as it becomes a source of versatile emergent electromagnetic responses~\cite{Ohgushi_PhysRevB.62.R6065,Bruno_PhysRevLett.93.096806,nagaosa2013topological,batista2016frustration,fujishiro2020engineering,hayami2021topological}. 
A magnetic skyrmion is a prototype of such topological spin textures~\cite{skyrme1962unified,Bogdanov89,Bogdanov94}.
Although the skyrmion crystal (SkX) has been originally observed in noncentrosymmetric magnets~\cite{Muhlbauer_2009skyrmion,yu2010real} where the Dzyaloshinskii-Moriya (DM) interaction plays an important role~\cite{dzyaloshinsky1958thermodynamic,moriya1960anisotropic,rossler2006spontaneous,Yi_PhysRevB.80.054416}, it has been recently observed in centrosymmetric magnets~\cite{kurumaji2019skyrmion,hirschberger2019skyrmion,khanh2020nanometric} by the frustrated exchange interactions~\cite{Okubo_PhysRevLett.108.017206,leonov2015multiply,Lin_PhysRevB.93.064430,Hayami_PhysRevB.93.184413} and/or the multiple-spin interactions in itinerant magnets~\cite{Ozawa_PhysRevLett.118.147205,Hayami_PhysRevB.95.224424,Hayami_PhysRevB.99.094420}. 
The essential feature by the different origins appears in the magnetic modulation period in the SkXs: The SkXs by the former mechanism usually lead to much longer magnetic periods than those by the latter ones~\cite{Tokura_doi:10.1021/acs.chemrev.0c00297}. 
Especially, a small skyrmion induces a large emergent magnetic field, and hence engineering the small skyrmion is promising for future spintronics applications, since it has great advantage to energy-efficient devices based on high density topological objects~\cite{Zhang_2020}.

The cubic chiral antiferromagnet EuPtSi is a typical compound to host such a SkX with the small magnetic period~\cite{kakihana2018giant,kaneko2019unique,tabata2019magnetic,kakihana2019unique}. 
The lattice structure of EuPtSi with the space group $P2_13$~\cite{ADROJA1990375,kakihana2017unique} is common to that of B20 compounds like MnSi~\cite{Muhlbauer_2009skyrmion}, and as a result, both compounds show similar temperature-field phase diagrams~\cite{kakihana2018giant}. 
Nevertheless, the magnetic periods of the SkXs are quite different with each other, e.g., $1.8$~nm for EuPtSi~\cite{kaneko2019unique,tabata2019magnetic} and $18$~nm for MnSi~\cite{ishikawa1984magnetic}, which results in giant topological Hall resistivity 0.12~$\mathrm{\mu\Omega \ cm}$ in EuPtSi~\cite{kakihana2018giant} compared to 
4.5 $\times 10^{-3}$$\mathrm{\mu \Omega \ cm}$ in MnSi~\cite{Neubauer_PhysRevLett.102.186602}. 
This implies that the short-period SkX in EuPtSi is not accounted for by the conventional DM-interaction-driven mechanism, which needs other factors as mentioned above. 
Moreover, there is another difference in a magnetic-field angle dependence.
The magnetism and conductivity in EuPtSi are sensitive to the magnetic-field directions~\cite{sakakibara2019fluctuation,Sakakibara_doi:10.7566/JPSJ.90.064701}; the signature of the SkX is observed when the magnetic field is in the [111] and [001] directions~\cite{takeuchi2019magnetic,kakihana2019unique,takeuchi2020angle,Sakakibara_doi:10.7566/JPSJ.90.064701}, while that in MnSi is observed irrespective of the magnetic-field direction~\cite{Bauer_PhysRevB.85.214418}.
In order to open up a way to the small skyrmion engineering for practical applications, it is important to understand the key ingredients of the short-period SkX in EuPtSi and to clarify the differences from MnSi. 

In the present study, we theoretically study the stability of the nanometric SkX with the EuPtSi in mind. 
We consider a minimal effective spin model with long-range Ruderman-Kittel-Kasuya-Yosida (RKKY)~\cite{Ruderman,Kasuya,Yosida1957}, biquadratic, and DM-type antisymmetric interactions that arise in itinerant magnets.
By performing simulated annealing, we find that our effective model well explains the mechanism of the short-period SkX and its behavior against the field direction.
There are two essences in the stabilization of the field-direction sensitive nanometric SkX: 
One is the low-symmetric ordering vector which brings about highly anisotropic responses against the field direction, and the other is the interplay between the long-range DM interaction arising from the spin-orbit coupling and the biquadratic interaction arising from the spin-charge coupling. 
We show that a fine balance between them determines the stability of the SkXs depending on the field directions. 
Our study will provide a reference not only to engineer the short-period SkX but also to explore further skyrmion-hosting materials in noncentrosymmetric itinerant magnets.

We consider a minimal effective spin model on a discrete cubic lattice including the effect of both spin-orbit and spin-charge couplings in noncentrosymmetric itinerant magnets.  
The Hamiltonian is given by 
\begin{align}
\mathcal{H} = \sum_{\nu=1}^{n}
&\left[-J\bm{S}_{\bm{Q}_\nu}\cdot\bm{S}_{-\bm{Q}_\nu}+\frac{K}{N}\left({\bm S}_{\bm{Q}_\nu}\cdot{\bm S}_{-\bm{Q}_\nu}\right)^2 \right.
\nonumber \\
&  \left.-i{\bm D}_\nu\cdot\left({\bm S}_{\bm{Q}_\nu}\times{\bm S}_{-\bm{Q}_{\nu}}\right)
\right]-\sum_{i}\bm{H}\cdot \bm{S}_i,
\label{eq:Ham}
\end{align}
where we take into account the interactions at particular wave vectors in momentum space $\{\bm{Q}_{\nu} \} 
= (\bm{Q}_1, \bm{Q}_2, \cdots, \bm{Q}_{n})$
($n$ is the number of the interaction channels) that arise from the Fermi surface instability in itinerant magnets. 
Here, $\bm{S}_{\bm{q}}$ is the Fourier transform of the localized spins $\bm{S}_i$ with $|\bm{S}_i|=1$, $\bm{S}_{\bm{q}}=(1/\sqrt{N})\sum_i \bm{S}_i e^{i\bm{q}\cdot \bm{r}_i}$ ($\bm{r}_i$ is the position vector at site $i$ and $N$ is the number of sites).
The model in Eq.~(\ref{eq:Ham}) consists of three long-range interactions derived from the perturbative analysis regarding the spin-charge coupling in the Kondo lattice model~\cite{Hayami_PhysRevB.95.224424,Hayami_PhysRevLett.121.137202}: the RKKY interaction with $J$ in the first term, the positive biquadratic interactions with $K>0$ in the second term, and the DM-type antisymmetric interaction with $\bm{D}_\nu \parallel \bm{Q}_\nu$ and $|\bm{D}_\nu|=D$ in the third term. 
In the derivation, we neglect two contributions. The one is the symmetric anisotropic bilinear interaction originating from the higher-order contribution of the spin-orbit coupling compared to the DM interaction, where we assume the small spin-orbit coupling~\cite{Hayami_PhysRevLett.121.137202}. 
The other is the four-spin interactions different from $({\bm S}_{\bm{Q}_\nu}\cdot{\bm S}_{-\bm{Q}_\nu})^2$ by assuming the distinct peak structures of the bare susceptibility of itinerant electrons~\cite{Hayami_PhysRevB.95.224424}.
The last term is the Zeeman coupling to an external magnetic field $\bm{H}$ whose directions are taken at $\bm{H} \parallel [111]$, $[001]$ and $[110]$ with $|\bm{H}|=H$. 
For simplicity, we neglect the higher-order contributions appearing under the magnetic field, such as the three-spin interactions~\cite{Hayami_PhysRevB.95.224424}.
We take $J=1$ as the energy scale. 

There are two key ingredients in the model in Eq.~(\ref{eq:Ham}). 
One is that the model includes the effects of both spin-charge and spin-orbit couplings, whose interplay induces nanometric topological spin crystals, such as a square SkX on a square lattice~\cite{Hayami_PhysRevLett.121.137202} and a hedgehog crystal on a cubic lattice~\cite{Okumura_PhysRevB.101.144416,Shimizu_PhysRevB.103.054427}.  
As the dominant interaction channels arise at the wave vectors at which the Fermi surfaces are nested, the short-period SkX naturally appears depending on the band structure and electron density.  
Furthermore, it is shown below that a small long-range DM interaction is enough to stabilize the short-period SkX in contrast to the short-range one that requires an unrealistic large value unless the other (multiple-spin) interactions are taken into account.

\begin{figure}[t!]
\begin{center}
\includegraphics[width=1.0\hsize]{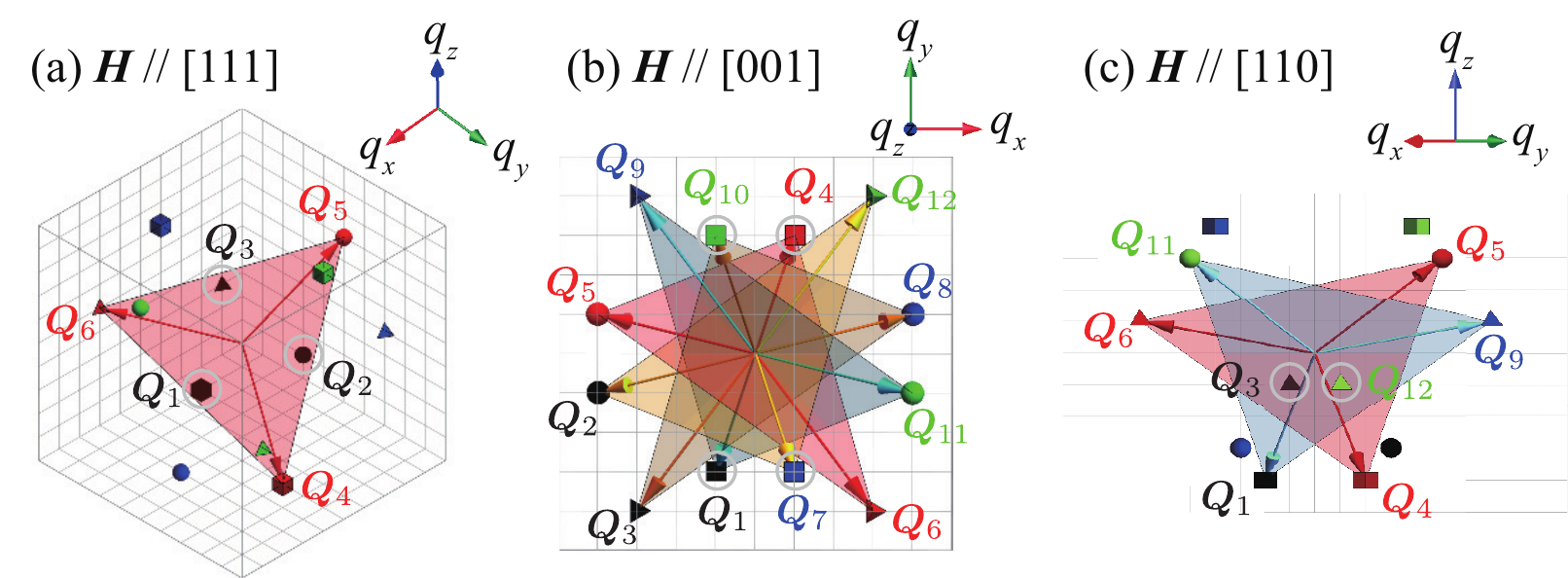} 
\caption{
\label{fig:Qvec}
(Color online) 
Positions of twelve ordering vectors $\bm{Q}_\nu$ ($\nu=1$-$12$) viewed from (a) [111], (b) [001], and (c) [110] directions. 
The open circles represent the ordering vectors for the single-$Q$ spiral state, while the triangles formed by the three arrows with the same color stand for the triple-$Q$ ordering vectors for the SkXs in each field direction. 
}
\end{center}
\end{figure}

The other is the symmetry of $\{\bm{Q}_{\nu} \}$. 
Although the dominant wave vectors are determined by the bare susceptibility depending on the electronic band structure and electron filling, we here suppose twelve low-symmetric ordering vectors $\{\bm{Q}_{\nu} \}$ with $n=12$, which are not lied on the high symmetry line and are connected by the symmetry operation under $P2_13$.
Specifically, we set $\bm{Q}_1=(-Q_a,-Q_b,-Q_c)$, $\bm{Q}_2=(-Q_c,-Q_a,-Q_b)$, $\bm{Q}_3=(-Q_b,-Q_c,-Q_a)$, $(\bm{Q}_{4},\bm{Q}_8, \bm{Q}_{12})=R^z(\bm{Q}_1, \bm{Q}_2, \bm{Q}_3)$, $(\bm{Q}_7, \bm{Q}_{11}, \bm{Q}_6)=R^y(\bm{Q}_1, \bm{Q}_2, \bm{Q}_3)$, and $(\bm{Q}_{10}, \bm{Q}_5, \bm{Q}_9)=R^x(\bm{Q}_1, \bm{Q}_2, \bm{Q}_3)$ with $Q_a=2\pi/15$, $Q_b=2\pi/5$, and $Q_c=8\pi/15$, where $R^{\alpha}$ ($\alpha=x,y,z$) represents $\pi$ rotation around the $\alpha$ axis. 
This mimics the situation of EuPtSi where the SkX under the [111] field has the triple-$\bm{q}$ vectors, $\bm{q}_1=(-\delta_3, \delta_1, \delta_2)$, $\bm{q}_2=(-\delta_2,\delta_3,-\delta_1)$, and $\bm{q}_3=(\delta_1,\delta_2,-\delta_3)$ for $\delta_1 \sim 0.09$, $\delta_2 \sim 0.20$, and $\delta_3 \sim 0.29$~\cite{kaneko2019unique,tabata2019magnetic}. 
The positions of $\{\bm{Q}_{\nu} \}$ are shown in Fig.~\ref{fig:Qvec}, where each $\{\bm{Q}_{\nu} \}$ is projected onto the (111) [Fig.~\ref{fig:Qvec}(a)], (001) [Fig.~\ref{fig:Qvec}(b)], and (110) [Fig.~\ref{fig:Qvec}(c)] planes.
These low-symmetric ordering vectors are important to induce a field-direction sensitive SkX, as shown below. 

Let us first study the low-temperature phase diagram of the model in Eq.~(\ref{eq:Ham}) at $K=0$ on the cubic lattice for $N=15^3$ by performing simulated annealing following the manner in Ref.~\citen{Hayami_PhysRevB.95.224424}~\cite{comment_size_EuPtSi}. 
The magnetic phases are identified by the magnetization $M_\alpha=(1/N)\sum_i S^\alpha_i$ and the magnetic moment with wave vector $\bm{q}$, $m^{\alpha}_{\bm{q}}=\sqrt{S^{\alpha}(\bm{q})/N}$ where $S^{\alpha}(\bm{q})$ is the $\alpha$ component of the spin structure factor $S^{\alpha}(\bm{q})=(1/N)\sum_{ij} S^\alpha_i S^\alpha_j e^{i \bm{q}\cdot (\bm{r}_i-\bm{r}_j)}$. 
We also calculate the spin scalar chirality $\chi^2=\chi^2_{xy}+\chi^2_{yz}+\chi^2_{zx}$ where $\chi_{\alpha\beta}=(1/N)\sum_{i\delta_\alpha \delta_\beta} \delta_{\alpha}\delta_{\beta} \bm{S}_i \cdot (\bm{S}_{i+\delta_{\alpha} \hat{\bm{x}}_\alpha}\times \bm{S}_{i+\delta_{\beta} \hat{\bm{x}}_\beta})$ for $\delta_{\alpha}=\pm 1$ ($\alpha=x,y,z$) and the $\alpha$-directional unit vector $\hat{\bm{x}}_\alpha$.

\begin{figure}[t!]
\begin{center}
\includegraphics[width=1.0\hsize]{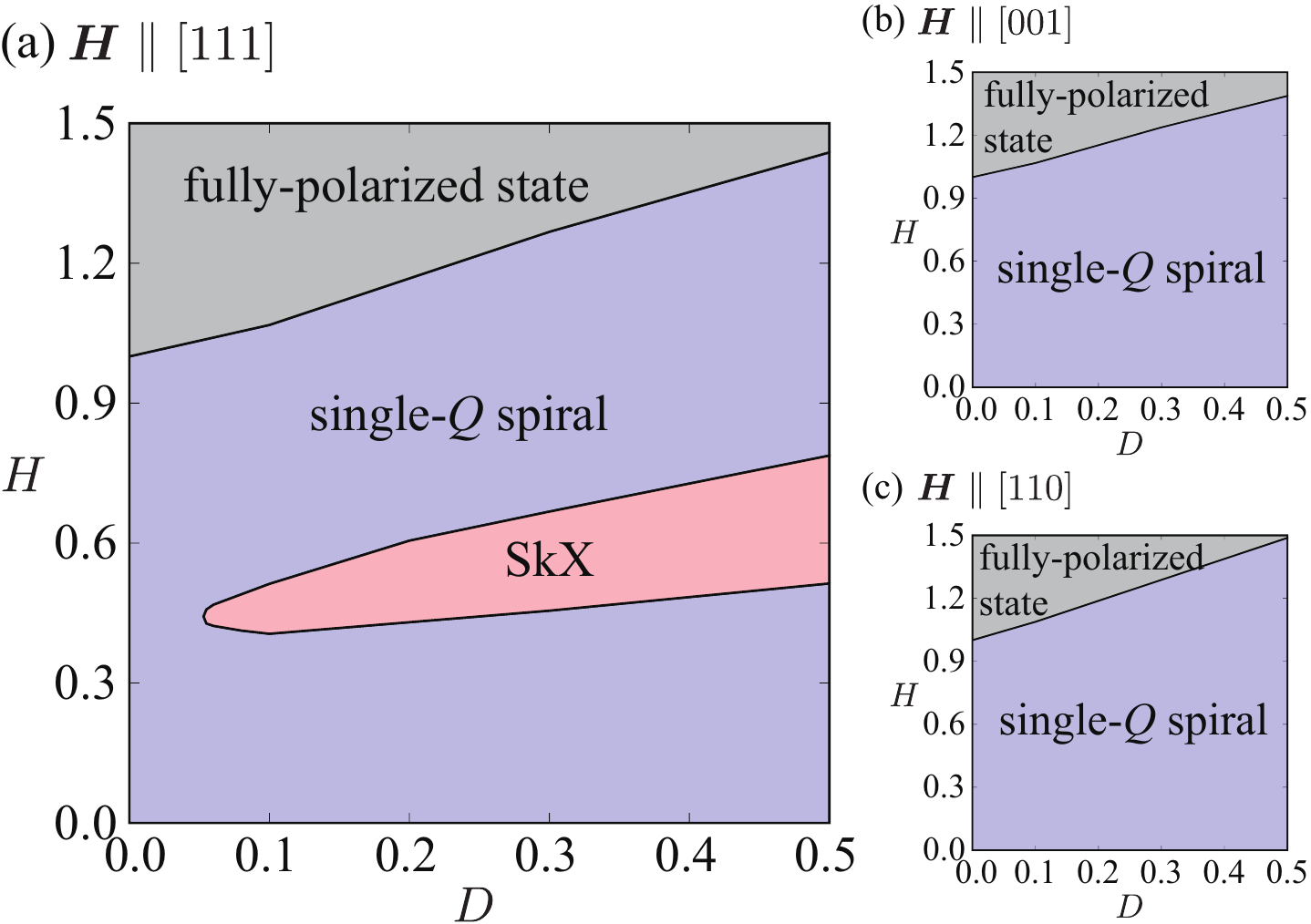} 
\caption{
\label{fig:PD_K=0}
(Color online) 
Magnetic phase diagram at a temperature of 0.01 obtained by the simulated annealing while varying the DM interaction $D$ and the external magnetic field $H$. 
The field directions are taken for the (a) [111], (b) [001], and (c) [110] directions. 
}
\end{center}
\end{figure}

\begin{figure}[t!]
\begin{center}
\includegraphics[width=1.0\hsize]{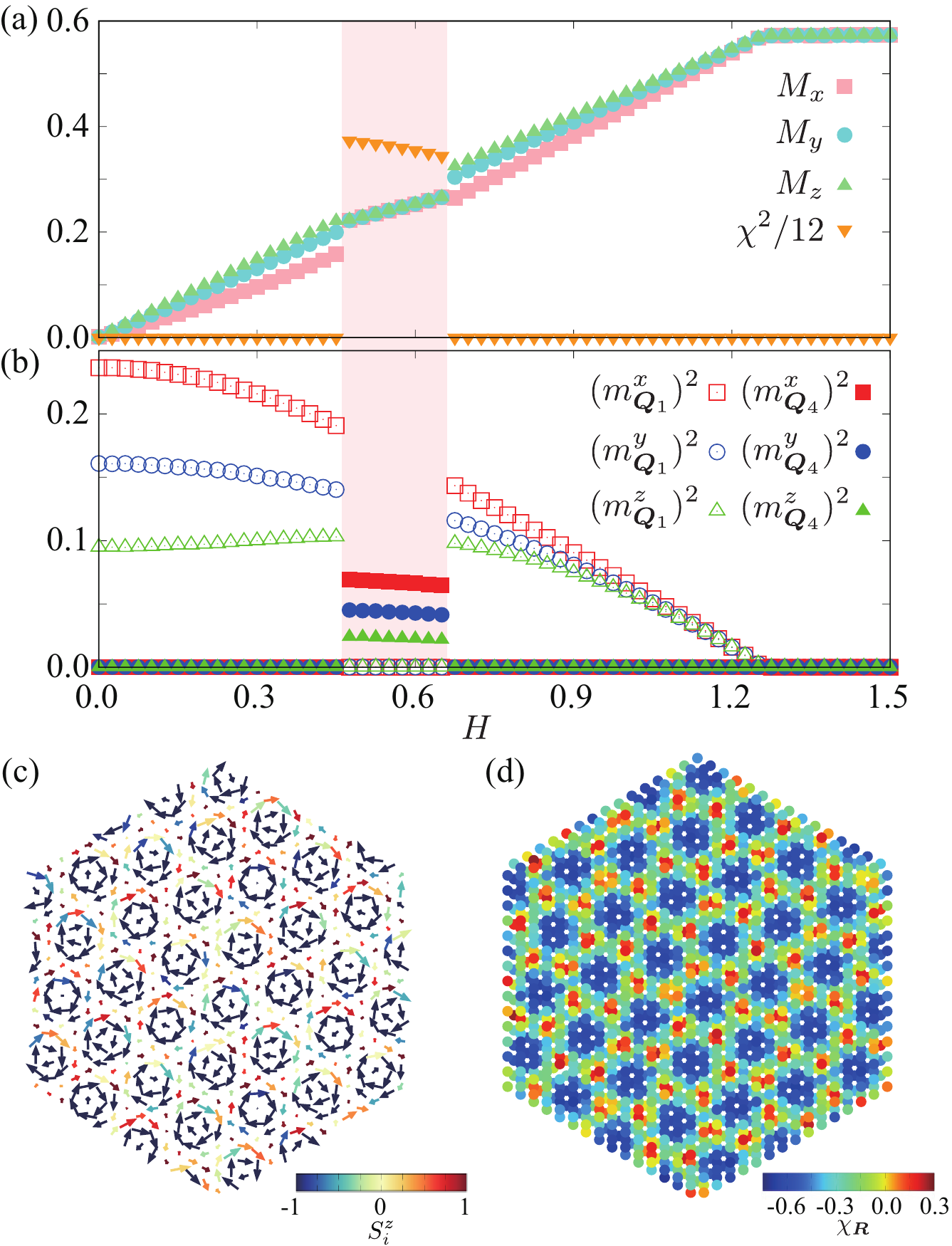} 
\caption{
\label{fig:Sq_111}
(Color online) 
$H$ dependences of (a) magnetization $M_{\alpha}$ and scalar chirality $\chi$ and (b) magnetic moments with relevant wave vectors $\bm{Q}_\nu$, $m^\alpha_{\bm{Q}_\nu}$, at $D=0.3$ and $K=0$ under the [111] field. 
Red regions in (a) and (b) represent the SkX phase. 
(c) Snapshot of the spin configuration in the SkX at $H=0.6$ viewed from the [111] direction. 
The arrow represents the averaged spin moment projected onto the (111) plane and the contour represents the parallel component to the (111) direction.
(d) Contour of the chirality corresponding to (c). 
}
\end{center}
\end{figure}

Figure~\ref{fig:PD_K=0}(a) displays the magnetic phase diagram under $\bm{H} \parallel [111]$ at $K=0$ while varying $D$ and $H$. 
Although a large portion of the phase diagram is occupied by the single-$Q$ spiral state, the SkX emerges for nonzero $D$ in the intermediate $H$ region; the SkX is stabilized for $D \gtrsim 0.05$ and the region extends while increasing $D$. 
We show $H$ dependences of the spin and chirality related quantities at $D=0.3$ in Figs.~\ref{fig:Sq_111}(a) and \ref{fig:Sq_111}(b). 
At $H=0$, the single-$Q$ spiral state is stabilized where the spiral plane is perpendicular to $\bm{Q}_{\nu}$ owing to the DM interaction. 
When applying $H$, the single-$Q$ spiral states with $\bm{Q}_1$, $\bm{Q}_2$, or $\bm{Q}_3$, which are connected by the threefold rotational symmetry around the [111] axis [Fig.~\ref{fig:Qvec}(a)], are chosen among twelve $\bm{Q}_{\nu}$ so as to maximize the Zeeman energy gain to favor the conical spiral perpendicular to $\bm{H}$. 
Accordingly, the spiral plane is tilted from $\bm{Q}_1$ (or $\bm{Q}_2$, $\bm{Q}_3$) to [111] direction while increasing $H$, as shown in Fig.~\ref{fig:Sq_111}(b). 

While further increasing $H$, $M_\alpha$ jumps and $\chi$ becomes nonzero, as shown in Fig.~\ref{fig:Sq_111}(a). 
This state is characterized by the triple-$Q$ 
 state whose threefold-symmetric ordering vectors $\bm{Q}_4$, $\bm{Q}_5$, and $\bm{Q}_6$ on the same (111) plane show equal intensities [see also Fig.~\ref{fig:Qvec}(a)]: $(m^x_{\bm{Q}_4})^2=(m^y_{\bm{Q}_5})^2=(m^z_{\bm{Q}_6})^2$, $(m^y_{\bm{Q}_4})^2=(m^z_{\bm{Q}_5})^2=(m^x_{\bm{Q}_6})^2$, and $(m^z_{\bm{Q}_4})^2=(m^x_{\bm{Q}_5})^2=(m^y_{\bm{Q}_6})^2$. 
These features signal the emergence of the SkX. 
Indeed, the skyrmion cores form a triangular crystal projected onto the (111) plane, as shown in Fig.~\ref{fig:Sq_111}(c). 
Simultaneously, the local chirality at the center of the square plaquette $\bm{R}$, $\chi_{\bm{R}}$, also forms the triangular crystal, resulting in a net scalar chirality as expected for the SkX. 
We confirm that this state has an integer skyrmion number.
The increment of $H$ in the SkX leads to the first-order phase transition to the single-$Q$ spiral state again, and continuously turns it into the fully-polarized state.

The results indicate that the SkX is stabilized by considering $D$ irrespective of $K$ under the [111] field. 
Notably, the SkX appears for a relatively small $D$, e.g., $D/J \simeq 0.05$, which means that the present model does not require a large DM interaction in contrast to the conventional spin model with the short-range interactions~\cite{comment_skyrmion_period}. 
Furthermore, we find that the emergent SkX is sensitive to the magnetic-field directions; no SkX appears under the $[001]$ and $[110]$ field directions, as shown in Figs.~\ref{fig:Sq_111}(b) and \ref{fig:Sq_111}(c), respectively. 

\begin{figure}[t!]
\begin{center}
\includegraphics[width=1.0\hsize]{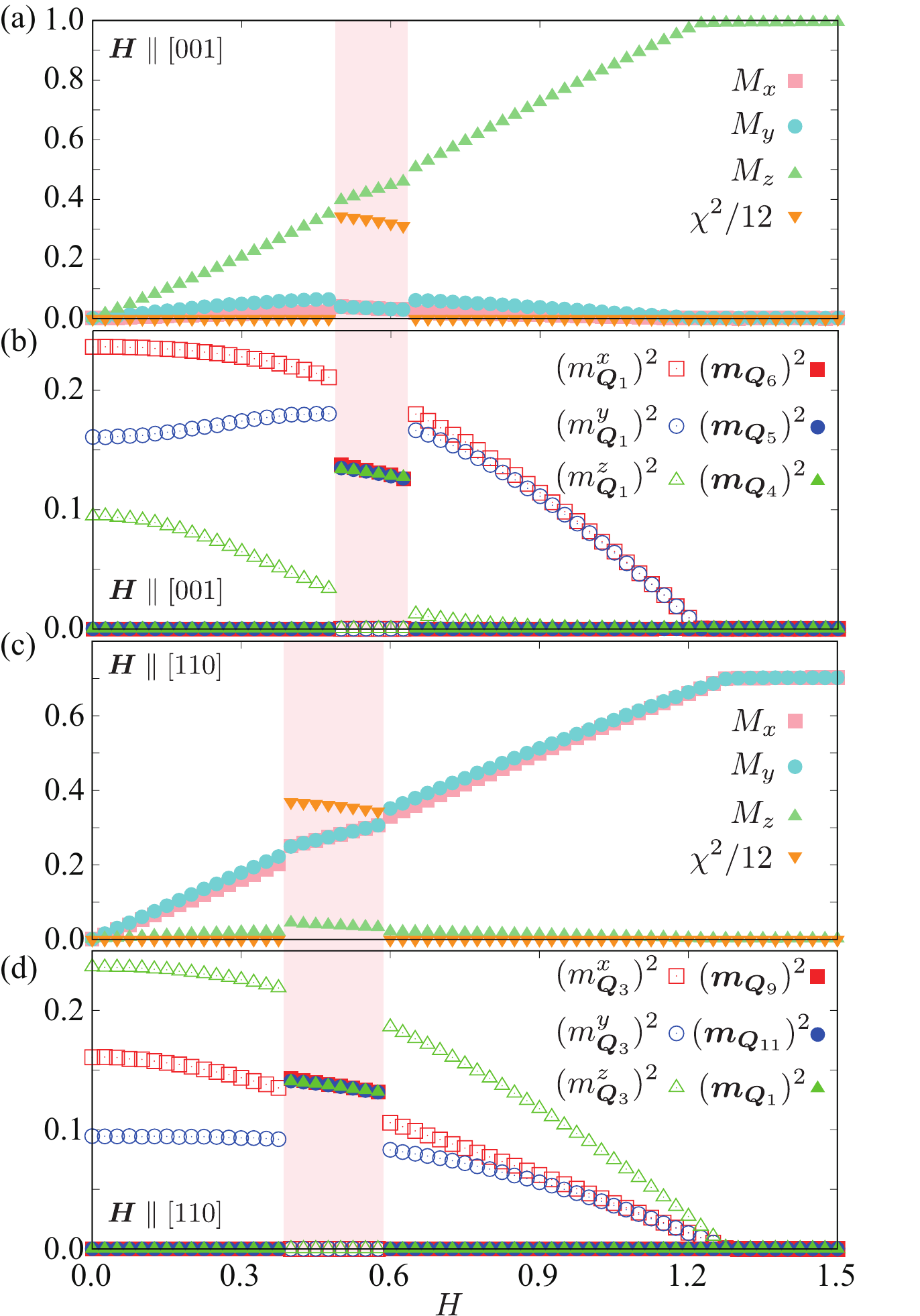} 
\caption{
\label{fig:Sq_001_110}
(Color online) 
$H$ dependences of (a,c) $M_\alpha$ and $\chi$ and (b,d) relevant $m^\alpha_{\bm{Q}_\nu}$ at $D=0.3$ for (a, b) $K=0.05$ under the [001] field and (c,d) $K=0.08$ under the [110] field. 
}
\end{center}
\end{figure}

Next, we consider the effect of $K$, which arises from the spin-charge coupling in itinerant magnets and tends to enhance the instability toward the SkX~\cite{Hayami_PhysRevB.95.224424}. 
We find that the introduction of $K$ can induce the SkXs even for $\bm{H}\parallel [001]$ and $\bm{H} \parallel [110]$. 
Figures~\ref{fig:Sq_001_110}(a) and \ref{fig:Sq_001_110}(b) [Figs.~\ref{fig:Sq_001_110}(c) and \ref{fig:Sq_001_110}(d)] show the $H$ dependences of $M_\alpha$, $\chi$, and $m^\alpha_{\bm{Q}_\eta}$ for $\bm{H}\parallel [001]$ ($\bm{H} \parallel [110]$) at $K=0.05$ ($K=0.08$). 
Both results show a similar phase sequence to that under $\bm{H} \parallel [111]$ in Figs.~\ref{fig:Sq_111}(a) and \ref{fig:Sq_111}(b); the single-$Q$ spiral state in the low-field region, the triple-$Q$ SkX in the intermediate region, and again the single-$Q$ spiral state in the high-field region before entering the fully-polarized state. 

Meanwhile, the nature of the single-$Q$ spiral states and the SkXs under each field direction is qualitatively different in terms of degeneracy in $\{\bm{Q}_\nu \}$. 
For $\bm{H}\parallel [001]$, there are four degenerate spirals with $\bm{Q}_1$, $\bm{Q}_4$, $\bm{Q}_7$, and $\bm{Q}_{10}$ and four degenerate SkXs with $(\bm{Q}_4, \bm{Q}_5, \bm{Q}_6)$, $(\bm{Q}_1, \bm{Q}_9, \bm{Q}_{11})$, $(\bm{Q}_2, \bm{Q}_7, \bm{Q}_{12})$, and $(\bm{Q}_3, \bm{Q}_8, \bm{Q}_{10})$, as shown in Fig.~\ref{fig:Qvec}(b). 
For $\bm{H}\parallel [110]$, there are two degenerate spirals with $\bm{Q}_3$ and $\bm{Q}_{12}$ and two degenerate SkXs with $(\bm{Q}_4, \bm{Q}_5, \bm{Q}_6)$ and $(\bm{Q}_1, \bm{Q}_9, \bm{Q}_{11})$, as shown in Fig.~\ref{fig:Qvec}(c). 
The $\bm{q}$ vectors in the single-$Q$ spirals are selected so as to maximize the parallel component to the magnetic field, while the triple-$Q$ vectors in the SkXs are chosen to satisfy $\bm{Q}_{\eta}+\bm{Q}_{\eta'}+\bm{Q}_{\eta''}=\bm{0}$ ($\eta,\eta',\eta''=1$-$12$) and to form the plane perpendicular to the field direction. 
Although the SkX in the [111] field satisfies the above conditions, those in the [001] and [110] fields do not fully satisfy the latter condition; the plane by the constituent triple-$Q$ vectors is tilted to the fields. 
As a result, the magnetization perpendicular to the fields is induced [Figs.~\ref{fig:Sq_001_110}(a) and \ref{fig:Sq_001_110}(c)], which leads to the Zeeman energy cost compared to the [111] field.

\begin{figure}[t!]
\begin{center}
\includegraphics[width=0.8\hsize]{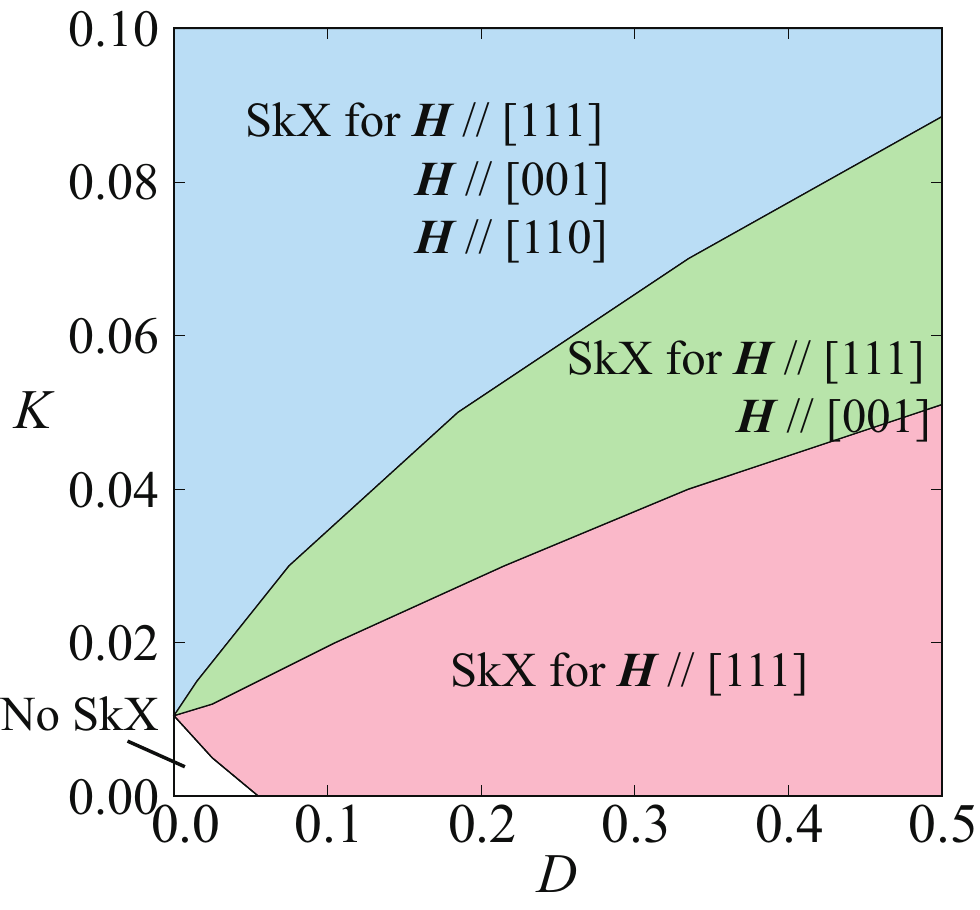} 
\caption{
\label{fig:PD_SkX}
(Color online) 
Magnetic phase diagram where the system undergoes a phase transition to the SkX in an applied magnetic field along the [111], [001], and [110] directions. 
``No SkX" represents that the SkX is not stabilized in all the field directions, while the other regions exhibit the SkX under the magnetic field along the directions denoted in the figure. 
}
\end{center}
\end{figure}

Thus, the critical values of $D$ and $K$ for the SkXs in each field direction are different. 
Figure~\ref{fig:PD_SkX} shows the phase diagram to denote the appearance of the SkXs for $\bm{H} \parallel [111]$, $[001]$, and $[110]$. 
Although the SkX for $\bm{H}\parallel [111]$ appears in almost all the regions including $K=0$ or $D=0$, while that for $\bm{H}\parallel [001]$ and $\bm{H}\parallel [110]$ emerges only for $K \neq 0$. 
The critical values of $K$ tend to be smaller for the [001] field than the [110] field, which might be attributed to the Zeeman energy loss of the single-$Q$ spiral state for $\bm{H}\parallel [001]$ compared to that for $\bm{H}\parallel [110]$.
In other words, the DM interaction tends to favor the SkXs in particular field directions, while the biquadratic interaction stabilizes the SkXs irrespective of field directions. 
This result provides useful information for understanding the key ingredients in field-direction sensitive SkXs observed in EuPtSi, as discussed below.

Finally, let us compare our results with the relevant material EuPtSi, which hosts the short-period SkXs depending on the magnetic-field direction; the topological Hall effect was observed along the [111] and [001] fields down to low temperatures, while they were not observed along the [110] field; especially, the SkX phase was identified in the [111] field~\cite{kakihana2018giant,kaneko2019unique,tabata2019magnetic,kakihana2019unique,takeuchi2019magnetic,takeuchi2020angle,Sakakibara_doi:10.7566/JPSJ.90.064701}.  
These anisotropic phase diagrams against the field direction are reproduced by the model parameters in the green region in Fig.~\ref{fig:PD_SkX}, which clearly suggests the importance of both spin-orbit and spin-charge couplings that might arise from the interplay between Eu-$4f$ localized spins and Pt-5$d$ itinerant electrons with the strong spin-orbit coupling. 
The model also captures that the SkX is surrounded by the single-$Q$ spiral state. 
These observations indicate that our effective spin model on the basis of the itinerant electron model well describes SkX physics in EuPtSi. 
One of the remaining issues is unidentified two successive phases (called A'-phase and B-phase) appearing in the [001] intermediate field~\cite{takeuchi2019magnetic,kakihana2019unique,takeuchi2020angle}, where the single SkX phase emerges in our results. 
Recalling that there are four degenerate SkXs in the [001] field, the degeneracy lifting by introducing an additional magnetic anisotropy and other multiple-spin interactions that are not taken into account in the model might be important to investigate the above two phases. 
In fact, it was shown that the magnetic anisotropy different from the DM interaction can also induce the SkXs and is relevant with the skyrmion-hosting materials~\cite{Hayami_PhysRevB.103.054422,yambe2021skyrmion}, such as GdRu$_2$Si$_2$~\cite{khanh2020nanometric,Hayami_PhysRevB.103.024439,Utesov_PhysRevB.103.064414,Wang_PhysRevB.103.104408} and Gd$_3$Ru$_4$Al$_{12}$~\cite{hirschberger2019skyrmion,Hirschberger_10.1088/1367-2630/abdef9}.
Thus, a further investigation to incorporate such additional magnetic anisotropy and multiple-spin interactions might be useful to clarify the origin of unidentified phases, which will be left for a future interesting issue.

To summarize, we have theoretically investigated the stability of the SkXs in chiral antiferromagnets on the basis of the minimal effective model phenomenologically taking into account long-range interactions characteristic of itinerant magnets. 
We showed that the interplay between the spin-orbit and spin-charge couplings stabilizes the nanometric SkX by performing simulated annealing. 
We also found that the spatial anisotropy that arises from the low-symmetric ordering vectors is enough to describe the field-direction sensitive SkXs. 
Our model well reproduced the SkXs observed in the $4f$-electron compound EuPtSi. 
Although more realistic model parameters based on the density-functional theory calculation might be desired for a further comparison, the effective spin model will be a good starting model to capture SkX physics in EuPtSi.
Moreover, as the effective model is universal in RKKY-driven itinerant magnets, the present results will be a useful reference to understand complex magnetic phases in other $f$-electron compounds, where the recent experiments imply the SkXs, such as EuPtGe~\cite{onuki2020unique} and EuAl$_4$~\cite{onuki2020unique,Shang_PhysRevB.103.L020405}.

\begin{acknowledgments}
This research was supported by JSPS KAKENHI Grants Numbers JP19K03752, JP19H01834, JP21H01037, and by JST PREST (JPMJPR20L8). 
Parts of the numerical calculations were performed in the supercomputing systems in ISSP, the University of Tokyo.
\end{acknowledgments}

\bibliographystyle{JPSJ}
\bibliography{ref}

\end{document}